\documentclass[12pt, draftcls, onecolumn, letterpaper]{IEEEtran}

\ifCLASSINFOpdf

\else

\fi

\ifCLASSOPTIONcompsoc
    \usepackage[caption=false, font=normalsize, labelfont=sf, textfont=sf]{subfig}
\else
    \usepackage[caption=false, font=normalsize]{subfig}
\fi
\usepackage{lipsum}%
\usepackage[dvipsnames]{xcolor}
\usepackage{algorithm,algorithmic}
\usepackage{balance}
\usepackage{multicol}   
\usepackage{cite}
\usepackage{gensymb}
\usepackage{multirow}
\usepackage{graphics}
\usepackage{epsfig}
\usepackage{graphicx}
\usepackage{epstopdf}
\usepackage{textcomp}
\usepackage{amsmath}
\usepackage{mathtools}
\usepackage{filecontents}
\usepackage{lipsum,color}
\usepackage{amssymb}
\usepackage{float}
\usepackage{colortbl} 
\usepackage{times} 
\usepackage{amsthm}  

\usepackage{amsfonts}

\theoremstyle{break}

\begin{document}
\title{A Study of Multihop mmW Aerial Backhaul Links
}

\author{\IEEEauthorblockN{ Mohammad~T.~Dabiri$^1$,~Mazen~O.~Hasna$^1$,~Tamer~Khattab$^1$,~and~Khalid~Qaraqe$^2$}
	\IEEEauthorblockA{\textit{$^1$Department of Electrical Engineering,} 
		\textit{Qatar University}, Doha, Qatar, \\
		E-mails: (m.dabiri;hasna;tkhattab)@qu.edu.qa.} \\
	\IEEEauthorblockA{\textit{$^2$Department of Electrical and Computer Engineering,} 
		\textit{Texas A\&M University at Qatar}, Doha, Qatar, 
		E-mail: khalid.qaraqe@qatar.tamu.edu}
	\thanks{This publication was made possible by NPRP13S-0130-200200 from the Qatar National Research Fund (a member of The Qatar Foundation). The statements made herein are solely the responsibility of the author[s].}
}

\maketitle
\vspace{-1cm}
\begin{abstract}
The main contribution of this paper is to analyze a long networked flying platform (NFP)-based millimeter wave (mmWave) backhaul link that is offered as a cost effective and easy to deploy solution  to connect a disaster or remote area to the nearest core network. For this aim, we characterize the backhaul channel as a function of realistic physical parameters such as heights and distances of obstacles along the route, flight altitude and the intensity of NFPs' vibrations, the real 3D antenna pattern provided by 3GPP, etc. For the characterized channel, we derive an analytical closed-form expression for the outage probability. Finally, using the obtained results, we provide a fast algorithm for the optimal parameter design of the considered system that minimizes the cost.
\end{abstract}
\IEEEpeerreviewmaketitle

\section{Introduction}

Tornadoes, hurricanes, earthquakes, tsunami and other major natural disasters all have the potential to cut or entirely destroy fibre infrastructure to the disaster area. 
Any disruption to the fragile fibre causes data outages which take days or weeks to locate and repair.
However, providing wireless alternative connectivity in the immediate moments after a disaster event is a key feature to facilitate rescue operations.
Implementing a long terrestrial wireless backhaul link faces major challenges, including providing a line of sight (LoS) between the nearest core network to the disaster area, especially for mountainous or forested areas.
Due to their unique capabilities such as flexibility, maneuverability,  and adaptive altitude adjustment, unmanned aerial vehicles (UAVs) acting as networked flying platforms (NFPs) can be considered as a promising solution to provide a temporary wireless backhaul connectivity while improving flexibility and reliability of backhaul operations \cite{alzenad2018fso,khawaja2019survey}.
More recently, millimeter wave (mmWave) backhauling has been proposed as a promising approach for aerial communications because of three reasons \cite{dabiri2019analytical}. First, unlike terrestrial mmWave communication links that suffer from blockage, the flying nature of UAVs offers a higher probability of LoS. Second, the large available bandwidth at mmWave frequencies can provide high data rate. Third, to compensate the negative effects of the high path-loss at the mmWave bands, the small wavelength enables the realization of a compact form of highly directive antenna arrays which is suitable for small UAVs with limited payload.
Although NFP-based mmWave backhaul link has been studied in recent works \cite{dabiri2019analytical,dabiri20203d,gapeyenko2018flexible,galkin2018backhaul, tafintsev2020aerial, 9411710, cicek2020backhaul, feng2018spectrum, yu2019uav}, the results of these studies are limited for short-distance NFP-based backhaul communications.



%
\begin{figure}
	\begin{center}
		\includegraphics[width=5 in]{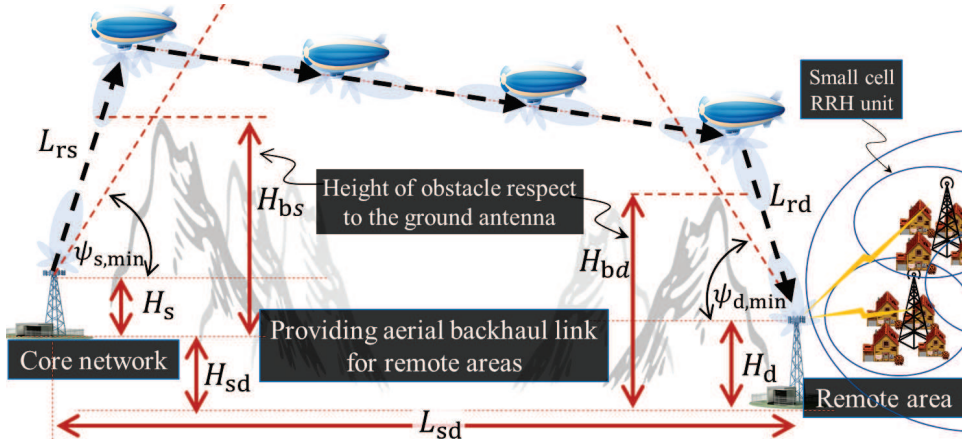}
		\caption{An illustration of a long aerial backhaul link in order to transfer data from the nearest core network to the disaster or remote area by using four NFPs. }
		\label{rn1}
	\end{center}
\end{figure}
%

In this study, we consider a long NFP-assisted backhaul link as shown in Fig. \ref{rn1} that is offered as a cost effective and easy to deploy solution  to connect a disaster or remote area to the nearest core network in a short time.
\footnote{The results of this study are provided for the types of NFPs that hover in the sky such as multi-rotor UAVs or tethered balloons and thus, for the optimal design of systems based on fixed-wing UAVs that are moving at high speed, a separate study should be done, which is beyond the scope of this work.}
In particular, we characterize a more practical scenario by tacking into account the effects of realistic physical parameters such as heights and distances of obstacles along the route, flight altitude and the severity of UAVs' vibrations, the real 3D antenna pattern provided by 3GPP, atmospheric channel loss as a function of temperature and air pressure, etc.
%
%
In this work, the QoS is evaluated using the outage probability metric, which is a very important criterion in wireless communication.
For this aim, we characterize the backhaul channel as a function of all aforementioned channel parameters and we derive an analytical closed-form expression for the outage probability.
By using Monte Carlo simulations, the accuracy of the derived analytical expression is verified. 
Then, we investigate the effects of key channel parameters such as antenna pattern gain, strength of UAV's vibrations, UAVs' positions in the sky, frequency bands, and physical parameters of a given region on the performance of the considered long NFP-based mmWave backhaul link in terms of  outage probability.
By providing sufficient simulation results, we carefully study the relationships between these parameters in order to reduce the negative effect of UAV vibrations by adjusting the 3D antenna pattern  and at the same time, decrease channel loss by adjusting optimal positions of UAVs in the sky compared to the obstacles.
Finally, using the obtained results, we provide a fast algorithm for the optimal design of system parameters that minimizes the cost.

\section{The System Model}
As illustrated in Fig. \ref{rn1}, we consider a multi-hop NFP-based mmWave backhaul link to transfer data from the nearest core network (source) to the disaster area (destination) where the terrestrial infrastructure is damaged due to natural events. 
%
%
%
%
%
Let $M$ be the number of NFPs where $U_1$ and $U_M$ denote the first and last NFPs, respectively, and $U_i$ for $i\in\{2,...,M-1\}$ denotes intermediate relays as shown in Fig. \ref{rn1}.
The core network sends signals towards $U_1$ and after passing through the intermediate relays, finally by the last relay, $U_M$, the signals will be directed to the destination in the remote area.

Each NFP is equipped with two mmWave directional antennas, one for the transmitter and one for the receiver. Let $A_{t,i}$ and $A_{r,i}$ denote respectively the transmitter and receiver antennas mounted on $U_i$ for $i\in\{1,...,M\}$. Also, let $A_s$ and $A_d$ be the core network and destination antennas, respectively.
Using location information obtained from the GPS, $U_1$ adjusts the direction of $A_{r,1}$ and $A_{t,1}$ toward the $A_s$ and $A_{r,2}$, respectively.  Similarly, for $i\in\{2,...,M-1\}$, $U_i$ adjusts the direction of $A_{r,i}$ and $A_{t,i}$ toward the $A_{t,i-1}$ and $A_{r,i+1}$, respectively. $U_M$ adjusts the direction of $A_{r,M}$ and $A_{t,M}$ toward the $A_{t,M-1}$ and $A_{d}$, respectively.
%
Assuming that each link is assigned a separate frequency band, the received signal at the first NFP, $i$th NFPs for $i\in\{2,...,M\}$, and destination  are obtained respectively as
\begin{align}
	\label{r1}
	\left\{
	\begin{array}{rl}
		&\!\!\!\!\!\!\!\! P_{r,1}= P_{t,s}   h_{L_s} G_s(\theta_{tx,s},\theta_{ty,s}) 
		G_{r,1}(\theta_{rx,1},\theta_{ry,1}), \\
		&\!\!\!\!\!\!\!\! P_{r,i} = P_{t,i-1} h_{L_{i-1}} G_{t,i-1}(\theta_{tx,i-1},\theta_{ty,i-1}) \\
		&~~~~\times      G_{r,i}(\theta_{rx,i},\theta_{ry,i}),  \\
		&\!\!\!\!\!\!\!\! P_{r,d}= P_{t,M}   h_{L_{M}} G_{t,M}(\theta_{tx,M},\theta_{ty,M}) 
		G_{d}(\theta_{rx,d},\theta_{ry,d}) ,
	\end{array}   \right. 
\end{align}
where $P_{t,s}$ and $G_{s}(\theta_{tx,s},\theta_{ty,s})$ are respectively the transmitted power and the antenna pattern gain of ground core network transmitter, $P_{t,i}$ and $G_{t,i}(\theta_{tx,i},\theta_{ty,i})$ are respectively the transmitted power and the antenna pattern gain of $A_{t,i}$, $G_{r,i}(\theta_{rx,i},\theta_{ry,i})$ is the antenna pattern gain of $A_{r,i}$, and $G_{d}(\theta_{rx,d},\theta_{ry,d})$ is antenna pattern gain of the ground destination.
The parameters $h_{L_0}$, $h_{L_M}$, and $h_{L_i}$ for $i\in\{1,...,M-1\}$ respectively denote the channel loss of source to $U_1$ link, $U_M$ to destination link, and $U_i$ to $U_{i+1}$ link. The UAVs' vibrations are characterized by random variables $\theta_{tx,i}$s, $\theta_{ty,i}$s, $\theta_{rx,i}$s, $\theta_{ry,i}$s.

%
As shown in Fig. \ref{rn1}, minimum elevation angels of the source is denoted by 
$\psi_{s,\text{min}}$ which is a function of source height $H_s$, and height of the nearest obstacle to the source $H_{bs}$. 
Also, minimum elevation angels of the destination is denoted by 
$\psi_{d,\text{min}}$ which is a function of destination height $H_d$, and height of the nearest obstacle to the destination $H_{bd}$. The height difference between the source and the destination is also denoted by $H_{sd}$. $L_{sd}$ is the horizon link length between the core network and the remote area and $H_{b,\text{max}}$ is the height of highest obstacle.

\subsubsection{Channel Propagation Loss}
In normal atmospheric conditions, water vapor (H$_2$O) and oxygen (O$_2$) molecules are strongly absorptive of radio signals, especially at mmWave frequencies and higher.
%
Channel loss is usually expressed in dB and it can be calculated using the formula
\begin{align}
	\label{f1}
	h_{L,\text{dB}}^{\text{tot}}(f_c) = 20\log\left(\frac{4 \pi L }{\lambda}\right)  + h_{L,\text{dB}}^{o,w}(f_c),
\end{align}
where $L$ is the link length (in m), $\lambda$ is the wavelength (in m), $f_c$ is mmWave frequency (in GHz),
$h_{L,\text{dB}}^{o,w}(f_c) = \frac{h_{L,\text{dB/km}}^{o,w}(f_c) L}{1000}$ is the attenuation due to oxygen and water (in dB), $h_{L,\text{dB/km}}^{o,w}(f_c)=h_{L,\text{dB/km}}^{o}(f_c)+h_{L,\text{dB/km}}^{w}(f_c)$ is the attenuation due to oxygen and water (in dB/km).
At 20°C surface temperature and at sea level, approximate expressions for the attenuation constants of oxygen and water vapor (in dB/km)  as defined by the International Telecommunications Union (ITU) are \cite{ITU_1}:
\begin{align}
	\label{po1}
	&h_{L,\text{dB/km}}^{o,0}(f_c) = 0.001\times f_c^2\\
	&\times \left\{
	\begin{array}{rl}
		\frac{6.09}{f_c^2+0.227} + \frac{4.81}{(f_c-57)^2+1.5} ~~~~~~~~~~~& ~~~  f_c<57  \\  
		h_{L,\text{dB/km}}^{o,0}(f_c=57) + 1.5(f_c-57)& ~~~  57<f_c<63 \\
		\frac{4.13}{(f_c-63)^2+1.1} + \frac{0.19}{(f_c-118.7)^2+2} ~~~~~~& ~~~  63<f_c<350
	\end{array} \right. \nonumber
\end{align}
and
\begin{align}
	\label{po2}
	&h_{L,\text{dB/km}}^{w,0}(f_c) = 0.0001\times f_c^2 \rho _0 \left(0.05  + \frac{3.6}{(f_c-22.2)^2+8.5}  \right. \nonumber \\
	& \left. + \frac{10.6}{(f_c-183.3)^2+9}    + \frac{8.9}{(f_c-325.4)^2+26.3} \right), ~~f_c<350,
\end{align}
where $\rho_0=7.5 ~\text{g/m}^3$ is the water vapor density at sea level, and $h_{L,\text{dB/km}}^{o,0}(f_c=57)$ is the value of the first expression at $f_c=57$.
In general, the attenuation constants of oxygen and water vapor are functions of altitude, since they depend on factors such as temperature and pressure. These quantities are often assumed to vary exponentially with height $H$, as $\rho(H) = \rho_0 \exp\left(-H/H_\text{scale}\right)$ where $H_\text{scale}$ is known as the scale height, which is typically 1-2 km.
From this, the specific attenuation as a function of height can be approximately modeled as
\begin{align}
	\label{po3}
	h_{L,\text{dB/km}}^{o,w}(f_c,H) = h_{L,\text{dB/km}}^{o,w,0}(f_c) \exp\left(-H/H_\text{scale}\right).
\end{align}
For a slant atmospheric path from height $H_1$ to $H_2$ at an angle $\psi$, the total atmospheric attenuation is determined by integration of the specific attenuation as
\begin{align}
	\label{po4}
	h_{L,\text{dB/km}}^{o,w}(f_c) \simeq 
	\frac{h_{L,\text{dB/km}}^{o,w,0}(f_c) \left( e^{-H_1/H_\text{scale}} - e^{-H_2/H_\text{scale}} \right) H_s}{\sin(\psi)}.
\end{align}
In our system model, both core network to first NFP (CU) link and last NFP node to destination (UD) link are slant and the inter-NFP links are approximately horizontal. 


\subsubsection{3D Antenna Pattern}
%
%
%
The array radiation gain is mainly formulated in the direction of $\theta$ and $\phi$. In our model, $\theta$ and $\phi$ can be defined as  functions of random variables (RVs) $\theta_{x}$ and $\theta_{y}$ as follows:
$\theta  = \tan^{-1}\left(\sqrt{\tan^2(\theta_{x})+\tan^2(\theta_{y})}\right)$,
$	\phi    =\tan^{-1}\left({\tan(\theta_{y})}\big/{\tan(\theta_{x})}\right)$. 
By taking into account the effect of all elements, the array radiation gain in the direction of angles $\theta_{x}$ and $\theta_{y}$ will be:
\begin{align}
	\label{p_1}
	G(\theta_{x},\theta_{y})  = G_0(N) \,
	\underbrace{G_e(\theta_{x},\theta_{y}) \,  G_a(\theta_{x},\theta_{y})}_{G'(\theta_{x},\theta_{y'})},
\end{align}
where $G_a$ is an array factor, $G_e$ is single element radiation pattern and $G_0$ is a constant defined in the sequel. 
From the 3GPP single element radiation pattern, $G_{e,\textrm{3dB}}=10\times\log_{10}(G_e)$ of each single antenna element is obtained as \cite{niu2015survey}. 

If the amplitude excitation of the entire array is uniform, then the array factor $G_a(\theta_{x},\theta_{y})$ for a square array of $N\times N$ elements can be obtained as \cite[eqs. (6.89) and (6.91)]{balanis2016antenna}
\begin{align}
	\label{f_1}
	&G_a(\theta_{x},\theta_{y}) = \\
	& \frac{\sin^2\left(\frac{N (k d_{x} \sin(\theta)\cos(\phi)+\beta_{x})}{2}\right)} 
	{N^2\sin^2\left(\frac{k d_{x} \sin(\theta)\cos(\phi)+\beta_{x}}{2}\right)}
	\frac{\sin^2\left(\frac{N (k d_{y} \sin(\theta)\sin(\phi)+\beta_{y})}{2}\right)} 
	{N^2\sin^2\left(\frac{k d_{y} \sin(\theta)\sin(\phi)+\beta_{y}}{2}\right)}\nonumber
\end{align}
where $\beta_{x}$ and $\beta_{y}$ are progressive phase shift between the elements along the  $x$ and $y$ axes, respectively. 
For a fair comparison between antennas with different $N$, we assume that the total radiated power of antennas with different $N$ are the same. From this, we have
\begin{align}
	\label{cv}
	G_0(N)=\left(   \int_0^{\pi}\int_0^{2\pi} G'(\theta,\phi) \sin(\theta) d\theta d\phi   \right)^{-1}.
\end{align}
More details on the elements and array radiation pattern is provided in \cite{niu2015survey,balanis2016antenna}.

%
\begin{figure}
	\begin{center}
		\includegraphics[width=3.5 in]{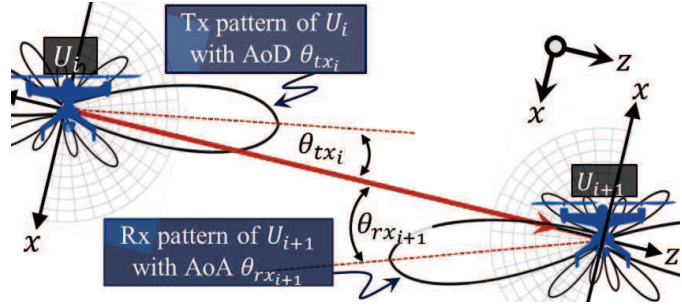}
		\caption{An illustration of UAVs' orientation fluctuations in the $x-z$ plane. }
		\label{rnd1}
	\end{center}
\end{figure}
%

\subsubsection{The Effect of UAV's Instabilities}
In practical situations, an error in the mechanical control system of UAVs, mechanical noise, position estimation errors, air pressure, and wind speed can affect the UAV's angular and position stability \cite{dabiri2018channel,dabiri2021uav}.
The instantaneous orientation of a UAV can randomly deviate from its means denoted by $\theta$. Let RVs 
$\theta_{tx,i}\sim \mathcal{N}(0,\sigma^2_{\theta})$ and 
$\theta_{rx,i}\sim \mathcal{N}(0,\sigma^2_{\theta})$ denote respectively the instantaneous angle of departure (AoD) of Tx and angle of arrival (AoA) of Rx antennas mounted on $U_i$ in the $x-z$ Cartesian coordinates as depicted in Fig. \ref{rnd1}, and RVs 
$\theta_{ty,i}\sim \mathcal{N}(0,\sigma^2_{\theta})$ and 
$\theta_{ry,i}\sim \mathcal{N}(0,\sigma^2_{\theta})$ denote the instantaneous AoD of Tx and AoA of Rx antennas mounted on $U_i$ in the $y-z$. 
%
%
%

\section{Optimal System Design}
For a given region with physical parameters such as air pressure, temperature, $\psi_{s,\text{min}}$, $\psi_{d,\text{min}}$, $H_s$, $H_{bs}$, $H_d$, $H_{bd}$, $H_{sd}$ $L_{sd}$, etc, our aim is to adjust the tunable system parameters such as NFPs' positions and antenna patterns in such a way that the cost is minimized.
Here, the cost mainly depends on the number of NFPs, denoted by $M$. 
From \eqref{f_1}, the antenna pattern is adjusted by the  number of antenna elements.
Let $N_{t,i}\times N_{t,i}$, $N_{r,i}\times N_{r,i}$, $N_{s}\times N_s$ and $N_{d} \times N_d$ denote the number of antenna elements of $A_{t,i}$, $A_{r,i}$, $A_{s}$, and $A_{d}$, respectively. 
The position of $U_1$ is characterized by $L_s$ and elevation angle $\psi_s$ where $\psi_{s,\text{min}}\leq \psi_s <\pi/2 $.
Moreover, the position of $U_M$ is characterized by $L_d$ and elevation angle $\psi_d$ where $\psi_{d,\text{min}}\leq \psi_d <\pi/2 $.
Our optimization problem can be formulated as follows:
\begin{subequations} \label{opt1} 
	\begin{IEEEeqnarray}{l} 
		\displaystyle \min_{ \substack{  
				\text{position of}~U_i,~i\in\{1,...,M\} \\ 
				N_{r,i},N_{t,i},~i\in\{1,...,M\} \\
				N_s,N_d     
		}  }  ~~~~~~~{M}\\
		~~~~~~~~~~~\textrm{s.t.}  ~~~\mathbb{P}_\text{out}< \mathbb{P}_\text{out,tar}~~~~~~~~~\\
		~~~~~~~~~~~~~~~~~\psi_{s,\text{min}}\leq \psi_s <\frac{\pi}{2}, \psi_{d,\text{min}}\leq \psi_d <\frac{\pi}{2},
	\end{IEEEeqnarray}
\end{subequations}
where $\mathbb{P}_\text{out}$ is the total outage probability which is calculated below, and $\mathbb{P}_\text{out,tar}$ is the target outage probability which is determined based on the requested QoS.
%

{\bf Lemma 1.}
{\it Outage probability of the considered system is derived as:}
\begin{align}
	\label{pou1}
	\mathbb{P}_\text{out} \simeq \mathbb{P}_\text{out,s1} + \mathbb{P}_\text{out,Md} + \sum_{i=2}^M \mathbb{P}_\text{out,i}
\end{align}
{\it where}
\begin{align}
	\label{vb2}
	&\mathbb{P}_\text{out,s1} = \sum_{j=1}^{J K} \left( e^{-\frac{2(j-1)^2}{J^2 N_{r,1}^2\sigma_\theta^2}} - e^{-\frac{2j^2}{J^2 N_{r,1}^2\sigma_\theta^2}}   \right)
	\mathbb{Y}(P_{r,\text{th}}-P'_{r,1}(j)), \nonumber \\
	&\mathbb{P}_\text{out,Md} = \sum_{j=1}^{J K} \left( e^{-\frac{2(j-1)^2}{J^2 N_{t,M}^2\sigma_\theta^2}} - e^{-\frac{2j^2}{J^2 N_{t,M}^2\sigma_\theta^2}}   \right)
	\mathbb{Y}(P_{r,\text{th}}-P'_{r,d}(j)),
\end{align}

\begin{align}
	\label{e4}
	&\mathbb{P}_\text{out,i} = \sum_{j=1}^{J K} \sum_{j'=1}^{J K} 
	\left( e^{-\frac{2(j-1)^2}{J^2 N_{t,i-1}^2\sigma_\theta^2}} - e^{-\frac{2j^2}{J^2 N_{t,i-1}^2\sigma_\theta^2}}   \right) \nonumber \\
	&~\times\left( e^{-\frac{2(j'-1)^2}{J^2 N_{r,i}^2\sigma_\theta^2}} - e^{-\frac{2j'^2}{J^2 N_{r,i}^2\sigma_\theta^2}}   \right)
	\mathbb{Y}(P_{r,\text{th}}-P'_{r,i}(j,j')),
\end{align}
{\it and $
	\mathbb{Y}(x)= \left\{
	\begin{array}{rl}
		1& ~~~ {\rm for}~~~ x\geq 0 \\
		0& ~~~ {\rm for}~~~ x< 0 \\
	\end{array} \right. 
	$ is the sign function, 
	$P'_{r,1}(j) = P_{t,s}   h_{L_s}(\psi_s,L_s) G_0(N_{r,1})G_0(N_s) N_s^2 10^{\frac{G_\text{max}}{5}} 
	\mathbb{G}(j,N_{q,i})$ and
	$P'_{r,i}(j,j') = P_{t,i-1} h_{L_{i-1}} G_0(N_{t,i-1}) G_0(N_{r,i}) 10^{\frac{G_\text{max}}{5}} 
	\mathbb{G}(j,N_{t,i-1}) \\ \mathbb{G}(j',N_{r,i}) 
	\mathbb{G}(j,N_{q,i})$.}
\begin{IEEEproof}
	Please refer to Appendix \ref{AppA}.
\end{IEEEproof}

Through the simulations, we will show that the results obtained from \eqref{pou1} are very close to the results obtained from the Monte-Carlo simulations.
More importantly, the runtime of \eqref{pou1} is much shorter than the Monte-Carlo simulation, especially for lower values of outage probability.

Without loss of generality and for notation simplicity, we assume that $H_{sd}\simeq 0$, $H_s<< H_{u1}$ and $H_d<<H_{ud}$. 
The effective value of CU link length is $L_{s,\text{ef}} = L_s \cos (\psi_s)$. Therefore $\psi_s$ must be decreased to increase $L_{s,\text{ef}}$.
On the other hand, based on \eqref{po4}, $\psi_s$ must be increased to reduce the attenuation loss. 
As a results, for any values of $L_s$, optimizing $\psi_s$ requires balancing an inherent tradeoff between decreasing $\psi_s$ to increase $L_{s,\text{ef}}$ and increasing it to decrease the attenuation loss. 
%
Since the transmitted power of ground station is higher than the transmitted power of NFP (due to the limitations of the NFP's transmitted power), the length of the CU link is greater than the UD link, so we expect $U_1$ to be at a higher altitude than $U_M$. 
Therefore, in order to provide LoS between $U_1$ and $U_M$, it is necessary that $U_1$ has a height higher than the highest obstacle, i.e., $H_{u_1}>H_{b,\text{max}}$.

{\bf Remark 1.} {\it   At 70 GHz, the optimal value for $\psi_s$ and 
	$\psi_d$ is placed in the following interval:}
\begin{align}
	\label{x1}
	H_\text{scale} <L_s \sin(\psi_w) < 2H_\text{scale},
\end{align}   
{where $w\in\{s,d\}$. Moreover, when $20^o<\psi_{q,\text{min}}<40^o$, the optimal value for $\psi_qw$ is close to $\psi_{w,\text{min}}$ and for $40^o<\psi_{w,\text{min}}$, the optimal value for $\psi_w$ is equal $\psi_{w,\text{min}}$.} 

{\bf Remark 2.} {\it  For $H_{u_i}>H_\text{scale}$, height changes in the order of a few hundred meters do not have a significant effect on the amount of channel attenuation at 70 GHz frequency.}

\begin{figure}
	\begin{center}
		\includegraphics[width=5 in]{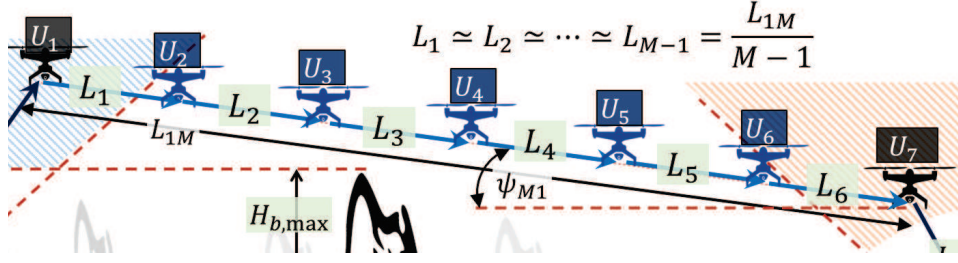}
		\caption{As illustrated for each given $M$, assuming that separate frequency bands are assigned to each link, the best place for NFPs is to be on a straight line. }
		\label{rn4}
	\end{center}
\end{figure}

In the next section, by providing simulations for conventional values of channel parameters at 70 GHz, we will confirm the results of Remarks 1 and 2. 
According to the results of Remark 2, for each given $M$, the best place for NFPs is to be on a straight line as shown in Fig. \ref{rn4}. Assuming that the NFPs are of the same type, the length of the links between the $U_i$s are equal and is calculated as follows:
\begin{align}
	\label{l1}
	L_i =  \frac{\sqrt{(L_{SD} -L_{s,\text{ef}}-L_{d,\text{ef}})^2 + (H_{u_1} - H_{u_M})^2}} 
	{ M-1},
\end{align}
where $i\in\{1,...,M-1\}$. Moreover, due to the symmetry, the optimal values for $N_{t,i}$s and $N_{r,i}$s must be the same as
\begin{align}
	\label{l2}
	N_{t,1} = ...=N_{t,M-1}=N_{r,2}=...=N_{r,M}=N_u.
\end{align}
%
%

As a summary of the aforementioned results, the tunable parameters of the considered long aerial backhaul link are listed as
\begin{align}\label{c1}
	\left\{
	\begin{array}{rl}
		&\!\!\!\!\!\text{CU link:}~~~~~~~~~~~~~ \psi_{s,\text{min}}<\psi_s<\pi/2,L_s,N_{r,1}, \\
		&\!\!\!\!\!\text{UD link:}~~~~~~~~~~~~~ \psi_{d,\text{min}}<\psi_d<\pi/2,L_d,N_{t,M}, \\
		&\!\!\!\!\!\text{Inter NFPs' links:}~~ L_i,N_u, M.
	\end{array}   \right. 
\end{align}
Equation \eqref{c1} is made up of nine variables, and it is clear that finding the optimal values for the nine variables simultaneously takes a lot of time.
Based on \eqref{l1}, $L_i$ is a function of tunable parameters $\psi_s$, $\psi_d$, $L_s$, $L_d$, and $M$, and once these parameters are known, the exact value of $L_i$ is calculated from \eqref{l1}. Moreover, based on the results of Remark 1, when $\psi_{s\text{max}}>20^o$ and $\psi_{d\text{max}}>20^o$, we can set $\psi_s\simeq \psi_{s\text{max}}$ and $\psi_d\simeq \psi_{d\text{max}}$. Therefore, the tunable parameters of our optimization problem can be reduced to six parameters as
\begin{align} \label{n3}
	\left\{L_s,N_{r,1},~~~ L_d,N_{t,M},~~~,N_u, M \right\}.
\end{align}

%
%

{\bf Lemma 2.} {\it For any given $\psi_s$ and $L_s$, the optimal value for $N_{r,1}$ can be obtained easily as follows:}
\begin{align} \label{opt2} 
	\displaystyle &\min_{ N_{r,1}} \left| C_1 h_{L_s}(\psi_s,L_s)  \left( \frac{\sin\left(\frac{N_{r,1} k d_{x} \sin(\frac{2 j}{J N_{r,1}})}{2}\right)} 
	{N_{r,1}\sin\left(\frac{k d_{x} \sin(\frac{2 j}{J N_{r,1}})}{2}\right)}
	\right)^2 - P_{r,\text{th}}\right|, \nonumber \\
	&\textrm{s.t.} ~~~ 1-\mathbb{P}_\text{out,tar} <  \sum_{j=1}^{J'} \left( e^{-\frac{2(j-1)^2}{J^2 N_{r,1}^2\sigma_\theta^2}} - e^{-\frac{2j^2}{J^2 N_{r,1}^2\sigma_\theta^2}}   \right).
\end{align}
{\it Similarly, for any given $\psi_d$ and $L_d$, we can compute the optimal value for $N_{t,M}$ from \eqref{opt2} by substituting the parameters 
	$N_{t,M}$, $N_d$, $h_{L_d}(\psi_d,L_d)$, and $P_{t,M}$ instead of the parameters 
	$N_{r,1}$, $N_s$, $h_{L_s}(\psi_s,L_s)$, and $P_{t,s}$, respectively.}
%
\begin{IEEEproof}
	Please refer to Appendix \ref{AppB}.
\end{IEEEproof}

Using the results of Lemma 2, it is observed that the 6-dimensional search space in \eqref{n3} is simplified to a 4-dimensional space containing variables $L_s$, $L_d$, $N_{u}$, and $M$, which significantly reduces the time to find the optimal parameters.
%
%

{From the obtained results, we provide Algorithm 1, which reduces the 8-dimensional search space to a 4-dimensional one, which significantly reduces the optimization time. 
	As discussed, for any given $\sigma_\theta$, the optimal values of $N_{r,1}$ and $N_{t,M}$, independent of the other tunable parameters,  are only a function of $L_s$ and $L_d$, respectively.  
	Therefore, in the first two {\bf for} loops of Algorithm 1, we calculate and store the optimal vectors of $\bar{N}_{r,1}$ and $\bar{N}_{t,M}$ based on the given $\sigma_\theta$ and intervals $L_{s,\text{min}}\leq L_s\leq L_{s,\text{max}}$, and  $L_{d,\text{min}}\leq L_d\leq L_{d,\text{max}}$.
	Then, each time we increase $M$ and calculate the optimal values for tunable parameters according to the three-dimensional {\bf for} loops.
	The first value of $M$ for which the end-to-end outage probability is lower than the threshold is the optimal value for $M$ and Algorithm 1 is terminated.
	In addition, to reduce the execution time of the algorithm, instead of the time consuming Monte Carlo simulations, the equation presented in Lemma 1 is used to calculate the outage probability.}

%
\begin{algorithm}[!htbp]
	\label{m1}
	\caption{Optimal design of parameters. }
	\textbf{Input}:	$\sigma_{to}$, $L_{sd}$, $\psi_\textrm{s,\text{min}}$, $\psi_\textrm{d,\text{min}}$, $f_c$, $P_{t,s}$, $P_{t,i}$, $N_s$, $N_d$, 
	$H_{b,\text{max}}$, $H_\text{scale}$, $H_{bd}$, $G_\text{max}$, $\mathbb{P}_\text{out,tr}$;\\ 
	\textbf{Output}: optimal pattern $\big(N_{r,1}^\text{opt},N_{t,M}^\text{opt}, N_i^\text{opt}\big)$, 
	optimal NFPs' positions  $(L_s^\text{opt}, L_d^\text{opt}, L_i^\text{opt})$, optimal number of NFPs $M_\text{opt}$;\\ 
	Initialize: $M=1$, $\mathbb{P}_\text{out,min}=1$;\\
	\textbf{for}:	$L_s=H_{b,\text{max}}/\sin({\psi_{s,\text{min}}})$ to $L_{s,\text{max}}$  \textbf{do} \\
	$|$\text{~~} find $N_{r,1}$ based on Lemma 2 and store in $\bar{N}_{r,1}$, \\
	\textbf{end for} \\
	\textbf{for}:	$L_d=H_{bd}/\sin({\psi_{d,\text{min}}})$ to $L_{d,\text{max}}$  \textbf{do} \\
	$|$\text{~~} find $N_{t,M}$ based on Lemma 2 and store in $\bar{N}_{t,M}$, \\
	\textbf{end for} \\
	\textbf{while}:  $\mathbb{P}_\text{out,min}>\mathbb{P}_\text{out,tr}$   \textbf{do} 
	$M = M+1$, \\
	$|$\text{~~} \textbf{for}:	$L_s=H_{b,\text{max}}/\sin({\psi_{s,\text{min}}})$ to $L_{s,\text{max}}$  \textbf{do} \\
	$|$\text{~~ ~~~} find $N_{r,1}$ from vector $\bar{N}_{r,1}$, \\
	$|$\text{~~ ~~~} \textbf{for}:	$L_d=H_{bd}/\sin({\psi_{d,\text{min}}})$ to $L_{d,\text{max}}$ \textbf{do}\\
	$|$\text{~~ ~~~ ~~~} find $N_{t,M}$ from vector $\bar{N}_{t,M}$,\\
	$|$\text{~~ ~~~ ~~~} compute $L_i$ based on \eqref{l1},\\
	$|$\text{~~ ~~~ ~~~} \textbf{for}:  $N_u=N_{u,\text{min}}$ to $N_{u,\text{max}}$ \textbf{do}\\
	$|$\text{~~ ~~~ ~~~ ~~~} compute $\mathbb{P}_\text{out}$ based on Lemma 1, \\
	$|$\text{~~ ~~~ ~~~ ~~~} \textbf{if}  $\mathbb{P}_\text{out}<\mathbb{P}_\text{out,min}$   \textbf{then do} $\mathbb{P}_\text{out,min} = \mathbb{P}_\text{out}$,\\ 
	$|$\text{~~ ~~~ ~~~ ~~~ ~~~} $N_{r,1}=N_{r,1}^\text{opt},N_{t,M}=N_{t,M}^\text{opt}, N_i=N_i^\text{opt}$, \\
	$|$\text{~~ ~~~ ~~~ ~~~ ~~~} $L_s=L_s^\text{opt}, L_d=L_d^\text{opt}, L_i=L_i^\text{opt}$, $M=M_\text{opt}$ \\
	$|$\text{~~ ~~~ ~~~ ~~~} \textbf{end if} \\    
	$|$\text{~~ ~~~ ~~~} \textbf{end for}\\
	$|$\text{~~ ~~~} \textbf{end for}\\
	$|$\text{~~} \textbf{end for} \\
	\textbf{end while}
\end{algorithm}
%

\section{Simulation Results}


%
\begin{figure}
	\begin{center}
		\includegraphics[width=4 in]{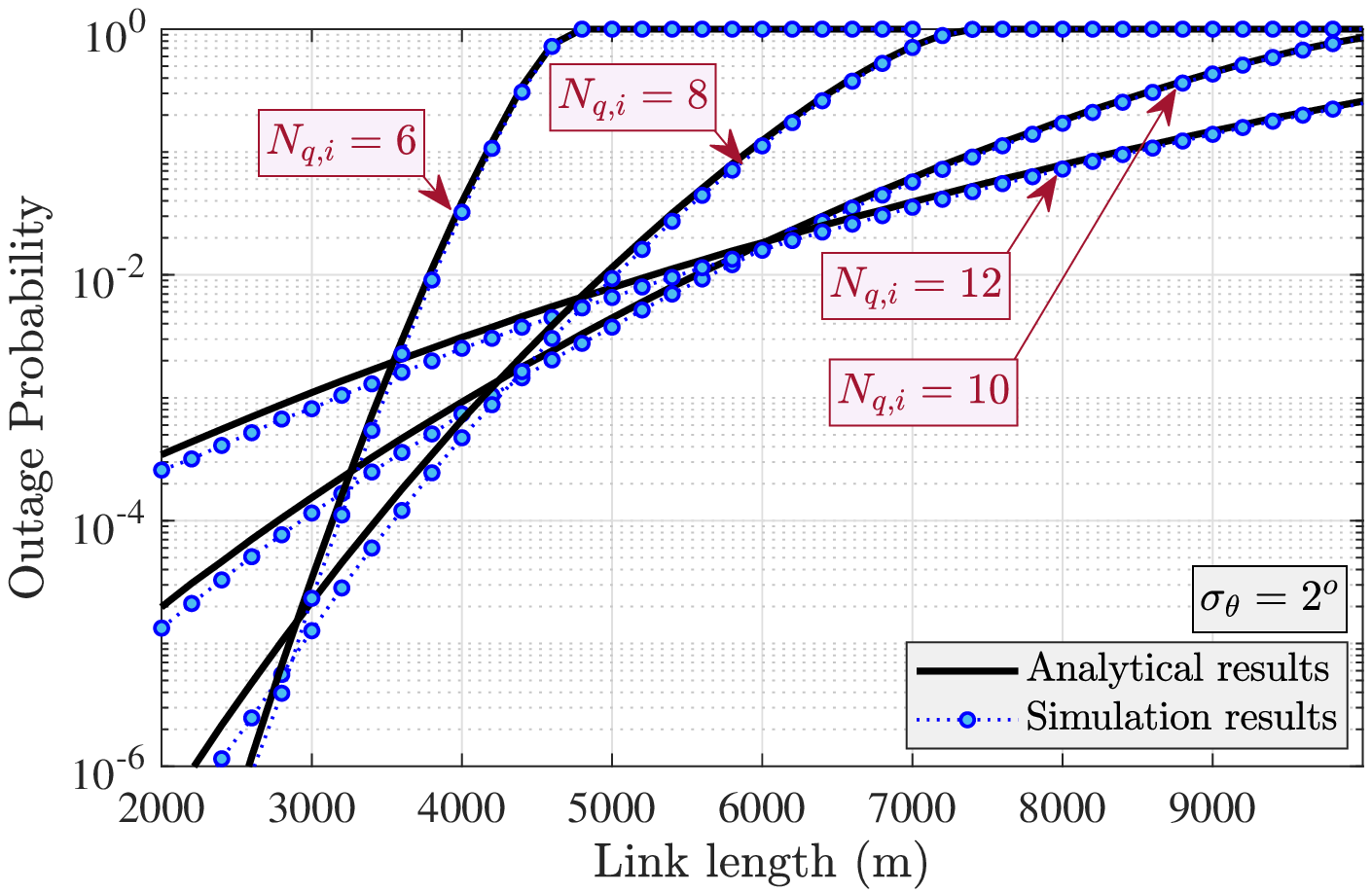}
		\caption{Outage probability of inter NFPs links versus link length for $H_{u_i}=2.5$ km, $f_c=70$ GHz, and $\sigma_\theta=2^o$.}
		\label{xb4}
	\end{center}
\end{figure}
%

In Fig. \ref{xb4}, outage probability of inter NFPs links is depicted versus link length for $f_c=70$ GHz and different antenna patterns characterized by $N_{q,i}$. The results of Fig. \ref{xb4} are obtained for $P_{t,i}=200$ mW, $\sigma_\theta=1.5^o$ and $\sigma_\theta=2^o$.  
For shorter link lengths, smaller values of $N_{q,i}$ that create a larger beam width perform better than larger ones. 
But as the link length increases, the channel loss increases and as a result, the average received signal power at the receiver antenna decreases. 
Therefore, we expect that the higher antenna pattern gain performs better for longer link lengths. 
The results of Fig. \ref{xb4} confirm the accuracy of this point. 

\begin{figure}
	\begin{center}
		\includegraphics[width=4 in]{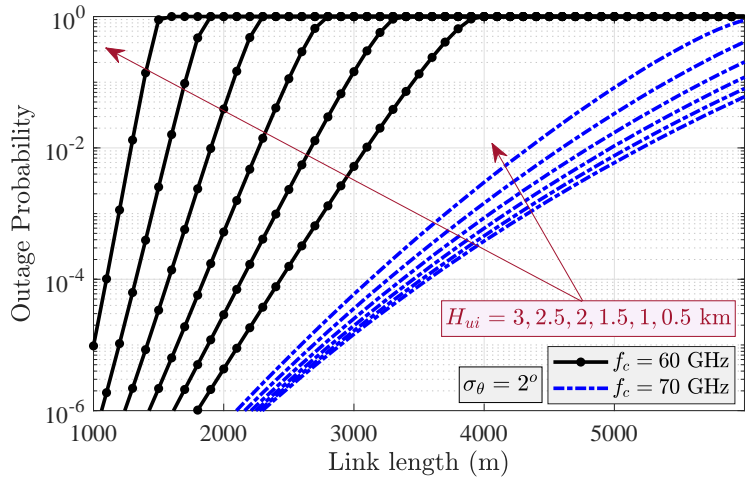}
		\caption{Comparison of the effect of $H_{u_i}$ on the outage probability of inter NFP links for frequency bands $f_c=60$ GHz and $f_c=70$ GHz.}
		\label{nd2}
	\end{center}
\end{figure}

Oxygen and water vapor can be strongly absorptive of radio signals, especially at mmWave frequencies and higher. 
The 60 GHz band is considered as unlicensed communication bands, however atmospheric attenuation peaks at 60 GHz, with a value of over 15 dB/km.
On the other hand, some frequency bands higher than 60 GHz such as 71–76, 81–86 and 92–95 GHz bands do not suffer from oxygen absorption, but require a transmitting license in the US from the Federal Communications Commission (FCC).
In Fig. \ref{nd2}, we plot the outage probability of inter-NFP links in Fig. \ref{nd2} for different NFP heights and two frequency bands $f_c=60$ and 70 GHz. 
For $f_c=60$ GHz, as the height of the NFPs increases, it is observed that the maximum achievable link length increases and thus,
this confirms that for the design of the considered system in the 60 GHz frequency band, the optimal choice of flight altitude of the NFPs is of great importance.
However, for $f_c=70$ GHz, it is observed that with increasing the altitude of the NFPs from 2 to 3 km, the performance of the system does not change significantly.

%
%

Finally, in Table \ref{I3}, for a given region with physical parameters $L_{sd}=40$ km, $\psi_{s,\text{min}}=40^o$, $\psi_{d,\text{min}}=20^o$, $H_{b\text{max}}=2$ km, $H_\text{scale}=1.5$ km, and $T=20^o$C,  we find the optimal values for the channel parameters such as optimal antenna pattern characterized by $\big(N_{r,1}^\text{opt},N_{t,M}^\text{opt}, N_i^\text{opt}\big)$, optimal NFPs' positions characterized by  $(L_s^\text{opt}, L_d^\text{opt}, L_i^\text{opt})$, optimal number of NFPs, denoted by $M_\text{opt}$. The results of Table \ref{I3} are obtained for two different NFPs' instabilities $\sigma_\theta=2^o$ and $\sigma_\theta=1.5^o$. From the results of Table \ref{I3}, for $\sigma_\theta=2^o$, the minimum required number of NFPs is $M=10$ to guarantee $\mathbb{P}_\text{out}<10^{-3}$ in the considered 40 km distance between the core network and the remote area, while for NFPs with higher stability $\sigma_\theta=1.5^o$, the minimum number of NFPs is reduced to $M=7$.

\begin{table}
	\caption{Optimal parameter design for $L_{sd}=40$ km, and two different NFPs' stability $\sigma_\theta=2^o$ and $1.5^o$.} 
	\centering 
	\begin{tabular}{| l ||  c | c | c | c | c | c | c |} 
		\hline 
		$\sigma_\theta$& 
		$N_{r,1}^\text{opt}$& $N_{t,M}^\text{opt}$&$N_i^\text{opt}$ &
		$L_s^\text{opt}$ &$L_d^\text{opt}$ & $L_i^\text{opt}$ &           $M_\text{opt}$ \\
		& 
		& & &
		(km) & (km) & (km) & \\ [.5ex] 
		\hline\hline 
		%
		$2^o$  & 6 & 6 & 8 &
		9.6  & 6.2  & 3.1  &
		10 \\ \hline
		$1.5^o$  & 8 & 8 & 10 &
		7.4  & 11.4  & 4.3  &
		7 \\
		\hline               
	\end{tabular}
	\label{I3} 
\end{table}



\appendices

\section{}
\label{AppA}
For the considered multihop DF relaying system, the outage probability of the end-to-end system is given by \cite{hasna2003outage}
\begin{align}
	\label{pon}
	\mathbb{P}_\text{out} &= \text{Prob}\left\{\text{min}(P_{r,1},...,P_{r,M},P_{r,d})<P_{r,\text{th}}\right\} \nonumber\\
	&= 1-(1-\mathbb{P}_\text{out,s1})(1-\mathbb{P}_\text{out,Md})\prod_{i=2}^M(1-\mathbb{P}_\text{out,i}),
\end{align}
where $\mathbb{P}_\text{out,s1}$, $\mathbb{P}_\text{out,Md}$, and $\mathbb{P}_\text{out,i}$ denote the outage probability of $A_s$ to $A_{r,1}$ link, $A_{t,M}$ to $A_d$ link, and $A_{t,i-1}$ to $A_{r,i}$ link, respectively,
and $P_{r,\text{th}}$ is the received power threshold.
For lower values of outage probability, \eqref{pon} can be approximated as \eqref{pou1}.

Considering the approximate symmetry of the antenna pattern along the $\psi$ axis, \eqref{p_1} can be approximated as
\begin{align}
	\label{f_a2}
	G(\theta_{q_i}) &=  G_0(N_{q,i}) 10^{\frac{G_\text{max}}{10}} 
	\left( \frac{\sin\left(\frac{N_{q,i} k d_{x} \sin(\theta_{qU_i})}{2}\right)} 
	{N_{q,i}\sin\left(\frac{k d_{x} \sin(\theta_{qU_i})}{2}\right)}
	\right)^2,
\end{align}
where $q\in{t,r}$, $\theta_{q_i}=\sqrt{\theta_{qx_i}^2 + \theta_{qy_i}^2}$. Since RVs $\theta_{qx_i}$ and $\theta_{qy_i}$ have Gaussian distributions, the random variable $\theta_{q_i}$ follows  the Rayleigh distribution as
\begin{align}
	\label{k1}
	f_{\theta_{q_i}}(\theta_{q_i}) = \frac{\theta_{q_i}}{\sigma_\theta^2}\exp\left(-\theta_{q_i}^2/2\sigma_\theta^2\right).
\end{align}
Let us approximate \eqref{f_a2} as 
\begin{align}
	\label{k2}
	&G(\theta_{q_i}) \simeq     G_0(N_{q,i})  10^{\frac{G_\text{max}}{10}} 
	\sum_{j=1}^{KJ} 
	\mathbb{G}(j,N_{q,i})
	\nonumber \\
	&~~~\times \left[ \mathbb{Y}\left(\theta_{q,i}- \frac{2 (j-1)}{J N_{q,i}} \right)  -  
	\mathbb{Y}\left(\theta_{q,i} - \frac{2 j}{J N_{q,i}} \right)  \right] ,
\end{align}
where 
$
\mathbb{G}(j,N_{q,i}) = \left( \frac{\sin\left(\frac{N_{q,i} k d_{x} \sin(\frac{2 j}{J N_{q,i}})}{2}\right)} 
{N_{q,i}\sin\left(\frac{k d_{x} \sin(\frac{2 j}{J N_{q,i}})}{2}\right)}
\right)^2
$,
$
\mathbb{Y}(x)= \left\{
\begin{array}{rl}
	1& ~~~ {\rm for}~~~ x\geq 0 \\
	0& ~~~ {\rm for}~~~ x< 0 \\
\end{array} \right. 
$ is the sign function,
and the parameters $J$ and $K$ are the natural numbers that for large values of $J$, \eqref{k2} tends to \eqref{f_a2}.
Also, $K=1$ refers to the main lobe of the antenna pattern and $K>1$ refers to the number of sidelobes.  
Note that the problem of orientation fluctuations is related to the antennas mounted on the NFPs. 
For the ground antennas $A_s$ and $A_d$, it is assumed that the ground station does not face weight and power limitations and uses a stabilizer with high accuracy and response speed to track the first and second drones. Based on this assumption, and using \eqref{r1}, \eqref{k1}, and \eqref{k2}, 
the outage probability of $A_s$ to $A_1$ link is obtained in Lemma 1.
From \eqref{r1} and \eqref{k2}, we have
\begin{align}  
	\label{e1}
	&P_{r,i} \simeq     P_{t,i-1} h_{L_{i-1}} G_0(N_{t,i-1}) G_0(N_{r,i}) 10^{\frac{G_\text{max}}{5}}  \sum_{j=1}^{KJ} \sum_{j'=1}^{KJ}
	\mathbb{G}(j,N_{t,i-1})\nonumber \\
	&
	\mathbb{G}(j',N_{r,i})
	\left[ \mathbb{Y}\left(\theta_{t,i-1}- \frac{2 (j-1)}{J N_{t,i-1}} \right) -   \right. 
	\left.  \mathbb{Y}\left(\theta_{t,i-1} - \frac{2 j}{J N_{t,i-1}} \right)  \right]\nonumber \\
	&  
	\left[ \mathbb{Y}\left(\theta_{r,i}- \frac{2 (j'-1)}{J N_{r,i}} \right)  -  
	\mathbb{Y}\left(\theta_{r,i} - \frac{2 j'}{J N_{r,i}} \right)  \right].  
\end{align}
As we see, $P_{r,i}$ is a function of two independent RVs $\theta_{t,i-1}$ and $\theta_{r,i}$. From \eqref{k1} and \eqref{e1} and using \cite{jeffrey2007table}, after some manipulations, the outage probability of $A_{t,i-1}$ to $A_{r,i}$ link is obtained in Lemma 1.

\section{}
\label{AppB}
Next, we show that for any given values of $L_w$ and $\psi_w$ where $w\in\{s,d\}$, $h_{L_w}$, we can obtain optimal values for $N_{r,1}$ and $N_{t,M}$. In the sequel, we prove this for $N_{r,1}$ and similarly we can prove this for $N_{t,M}$. 
For lower values of NFP's orientation fluctuations, we can approximate $P'_{r,1}(j)$ with a good accuracy as
\begin{align}
	\label{p1}
	&P'_{r,1}(j) \simeq \\
	&\left\{
	\begin{array}{rl}
		&\!\!\!\!\!\!\!C_1 h_{L_s}(\psi_s,L_s)  \left( \frac{\sin\left(\frac{N_{r,1} k d_{x} \sin(\frac{2 j}{J N_{r,1}})}{2}\right)} 
		{N_{r,1}\sin\left(\frac{k d_{x} \sin(\frac{2 j}{J N_{r,1}})}{2}\right)}
		\right)^2,~~~~j<J, \\
		&\!\!\!\!\!\!\!0, ~~~~~~~~~~~~~~~~~~~~~~~~~~~~~~~~~~~~~~~~~~~~~~~~~~~~~~j>0,
	\end{array} \right. \nonumber
\end{align}
where $C_1$ is a constant. Based on \eqref{p1}, it can easily be shown that $\frac{\text{d}P'_{r,1}(j)}{\text{d}j}\leq0$ and, therefore, $P'_{r,1}(j)$ is a descending function of $j$.
Using this and after some manipulations, we obtain
\begin{align}
	\label{vb5}
	&\mathbb{P}'_\text{out,s1}(N_{r,1})\simeq 1- \sum_{j=1}^{J'} \left( e^{-\frac{2(j-1)^2}{J^2 N_{r,1}^2\sigma_\theta^2}} - e^{-\frac{2j^2}{J^2 N_{r,1}^2\sigma_\theta^2}}   \right),
\end{align}
where $J'$ can be obtained easily from \eqref{p1} when $ P'_{r,1}(J') = P_{r,\text{th}}$. Based on these, for any given $h_{L_s}(\psi_s,L_s)$, independent of other tunable parameters, any $N_{r,1}$ that satisfies the conditions of \eqref{opt2} is the optimal value.


		


\end{document}